\documentclass[11pt,a4paper]{article}

\usepackage{jcappub}

\definecolor{darkorange}{rgb}{0.69,0.33,0.13}

\newcommand{\varphit}{\ensuremath{\tilde{\varphi}}}
\newcommand{\Phit}{\ensuremath{\tilde{\Phi}}}

\newcommand{\Rb}{\ensuremath{\bar{R}}}
\newcommand{\abar}{\ensuremath{\bar{a}}}
\newcommand{\chib}{\ensuremath{\bar{\chi}}}
\newcommand{\phib}{\ensuremath{\bar{\phi}}}
\newcommand{\rhob}{\ensuremath{\bar{\rho}}}
\newcommand{\Pb}{\ensuremath{\bar{P}}}

\newcommand{\RTA}{\ensuremath{R_{{\tt TA}}}}
\newcommand{\grad}{\ensuremath{ \vec{\nabla} }}
\newcommand{\Order}{\ensuremath{ {\cal O} }}

\newcommand{\LCDM}{{$\Lambda$CDM }}

\title{The maximum sizes of large scale structures in alternative theories of gravity}
\author[a,b]{Sourav Bhattacharya}
\author[c,d]{Konstantinos F.~Dialektopoulos}
\author[e]{Antonio Enea Romano}
\author[g,h]{Constantinos Skordis}
\author[f]{Theodore N. Tomaras}

\affiliation[a]{IUCAA, Pune University Campus, Ganeshkhind, Post Bag 4,  Pune 411 007, India}
\affiliation[b]{Department of Physics, Indian Institute of Technology  Ropar, Rupnagar, Punjab 140 001, India\footnote{Current affiliation}}
\affiliation[c]{Dipartimento di Fisica, Università di Napoli {}``Federico II'', Compl. Univ. di
Monte S. Angelo, Edificio G, Via Cinthia, I-80126, Napoli, Italy}
\affiliation[d]{INFN Sezione  di Napoli, Compl. Univ. di Monte S. Angelo, Edificio G, Via Cinthia, I-80126, Napoli, Italy.}
\affiliation[e]{Instituto de F\'{i}sica, Universidad de Antioquia, Calle 70 No. 52-21, Medell\'{i}n, Colombia.}
\affiliation[f]{ITCP and Department of Physics, University of Crete, 70013 Heraklion, Greece}
\affiliation[g]{Department of Physics, University of Cyprus, 1  Panepistimiou Street, Nicosia 2109, Cyprus.}
\affiliation[h]{Institute of Physics of the Czech Academy of Science, Na Slovance 2, Prague 8, Czech Republic.}

\emailAdd{sbhatta@iitrpr.ac.in} 
\emailAdd{kdialekt@gmail.com}
\emailAdd{aer@phys.ntu.edu.tw}
\emailAdd{skordis@ucy.ac.cy}
\emailAdd{tomaras@physics.uoc.gr}

\abstract{
The maximum size of a cosmic structure is given by the maximum turnaround radius -- the scale where the attraction due to its mass
 is balanced by the repulsion due to dark energy. We derive generic formulae for the estimation of the  maximum turnaround radius
in any theory of gravity obeying the Einstein equivalence principle, in two situations: on a spherically symmetric spacetime
 and on a perturbed Friedman-Robertson-Walker spacetime. We show that the two formulae agree.
As an application of our formula, we calculate the maximum turnaround radius  in the case of the Brans-Dicke theory of gravity.
We find that for this theory, such maximum sizes always 
lie above the \LCDM value, by a factor $1 + \frac{1}{3\omega}$, where $\omega\gg 1$ is the Brans-Dicke parameter, 
implying consistency of the theory with current data.
}

  
\begin{document}

\maketitle

\section{Introduction}
 The $\Lambda$-cold dark matter ($\Lambda$CDM) model is widely considered to be the simplest
 and most successful theoretical description of our universe, and finds support from 
a wide range of cosmological observations.
Despite its success, this model is  unfortunately  not without problems. While certain observational glitches have been reported
from time to time~\cite{Perivolaropoulos2008,FamaeyMcGaugh2011,SahniShafielooStarobinsky2014,BullEtAl2015},
the biggest challenge the \LCDM model has to face is the cosmological constant problem.~\footnote{The
cosmological constant problem is not manifest only for \LCDM but for any other theory with dynamical dark energy or
alternative to GR.}

The tiny value of the observed cosmological constant, $\Lambda \sim \Order(10^{-3} eV)^4$, 
that is needed for the model to be observationally viable, finds no compelling explanation from a quantum field theoretical point of view.
There had been numerous attempts to explain the value of $\Lambda$ by relating it to vacuum energy density of quantum fields,  
but all such attempts have either theoretical or observational inconsistencies~\cite{Weinberg1988,Martin2012}. A 
related problem is that de Sitter space may also be unstable to 
quantum corrections~\cite{AntoniadisIliopoulosTomaras1986,FloratosIliopoulosTomaras1987,Polyakov2012, Woodard2014}.

These conceptual and observational problems with the cosmological constant $\Lambda$ have triggered 
in recent years  vigorous research in alternatives to the \LCDM model. The chief agenda of these alternative models 
is to generate the effect of the dark energy through additional matter fields 
(for instance quintessence~\cite{CopelandSamiTsujikawa2006}), or, by replacing the theory of gravity on 
which \LCDM rests, i.e. General Relativity, by a different 
theory~\cite{CliftonEtAl2012, BullEtAl2015}(see also \cite{Padmanabhan2016} for a recent critique of the current status of cosmology). 
In order to discriminate between such alternative theories of gravity and GR, 
it is necessary to test all their possible observable consequences with cosmological observations. 
The next generation of cosmological surveys will offer a huge boost in precision making such tests 
possible~\cite{AbellEtAl2009,LaureijsEtAl2011,YahyaEtAl2014,SpergelEtAl2015,AghamousaEtAl2016}.

In this paper, we are particularly interested in one possible test of \LCDM and alternative theories of gravity, namely, 
the stability of the large scale cosmic structures~\cite{PavlidouTomaras2013}~\footnote{See also \cite{EingornKudinovaZhuk2012}
 for a different approach.}.
The maximum size of a large scale cosmic structure with a given mass $M$ 
 can be estimated using the maximum turnaround radius (or simply the turnaround radius $\RTA$ for short).
More precisely, $\RTA$ is the point where for radially moving test particles the attraction due to normal matter is balanced 
by the repulsion due to the dark energy. 
Specific theories are expected to lead to estimates for the turnaround radius, which depend on the theory parameters. 
If a certain theory predicts a maximum possible size smaller than the actual observed size of structures of mass $M$, 
the latter are expected to be unstable in the framework of that theory.
Thus, parameter ranges resulting to  maximum possible sizes smaller than what we observe are ruled out.

The turnaround radius was calculated in~\cite{Stuchlik1983,StuchlikHledik1999} in the wider context 
of geodesics of the Schwarzschild-de Sitter spacetime\footnote{The turnaround radius was called the ``static
  radius'' in ~\cite{Stuchlik1983,StuchlikHledik1999}.}. In a cosmological context,
the turnaround radius for spherical structures was calculated for $\Lambda$CDM in~\cite{PavlidouTomaras2013},
 and in ~\cite{PavlidouTetradisTomaras2014} for smooth dark energy. The turnaround radius as a cosmological observable 
was investigated in~\cite{TanoglidisPavlidouTomaras2014,TanoglidisPavlidouTomaras2016}.  In \cite{Lee2016} it was proposed to 
look for the violation of the maximum upper bound of $\RTA$ using the zero velocity surfaces 
of a large scale structure, by observing the peculiar velocity profiles of its members. 
It turns out that for structures as massive as $10^{15}M_{\odot}$ (e.g. the Virgo supercluster), the actual sizes 
lie very close and below the theoretical prediction of \LCDM~\cite{PavlidouTomaras2013}.
The structures studied in the references above are at sufficiently low redshifts ($z\sim 10^{-2}$),
and hence $\RTA$ measurements could provide a local indication and check for dark energy. In other words, it does not require 
any data coming from the high redshift Supernovae or from the early universe. 

Measuring the  turnaround radius offers yet another way of putting constraints on alternative gravity models.
For instance, the maximum turnaround radius has recently been calculated for 
a cubic galileon model~\cite{BhattacharyaDialektopoulosTomaras2015}. A method to calculate the 
turnaround radius in generic gravitational theories was
put foward in ~\cite{Faraoni2015,FaraoniLapierrePrain2015} by considering timelike geodesics, 
in the framework of alternative gravity theories admitting McVittie-like~\cite{McVittie1933,Gao2004,KaloperKlebanMartin2010} solutions.
We agree with the general formula for the turnaround, eq. 21, of ~\cite{Faraoni2015} but disagree in other results of that article 
 which appear to be in conflict with ours.~\footnote{In \cite{Faraoni2015} 
it is assumed that the potentials $\Phi$ and $\Psi$ have solutions $\sim \frac{G m}{r}$ under the assumption of spherical symmetry
(even in \LCDM), where the constant $m$ is the mass of the source. This however is incorrect as the correct solution (as can be 
verified by inspecting the McVittie solution) is  $\sim \frac{G m}{a r}$. Indeed, eq. 29 in \cite{Faraoni2015} gives a time-dependent 
turnaround radius, which is in disagreement with the known result for \LCDM. Our second source of disagreement is the recasting of the 
turnaround radius in terms of the areal radius. While we agree with the reasoning and with the relation between the comoving and areal radius,
 the contribution to the turnaround formula is of higher order in perturbation theory and should be neglected unless
higher order in perturbation theory solutions are also used.}

This article is organized as follows. In section \ref{sec_turnaround} we derive a general formula 
for the calculation of the turnaround radius, valid in any metric theory of gravity obeying the Einstein Equivalence Principle (EEP), 
a necessary assumption as the geodesic equation is used in our derivation. 
We perform our derivation in steps: (i) we first calculate the maximum turnaround radius in the case of  the \LCDM model using
static coordinates, (ii) we extend the static metric calculation to arbitrary theories (arbitrary static metrics), 
(iii) we re-calculate the turnaround radius using the McVittie metric and finally (iv) we relate the two types of calculation (static and McVittie)
 using cosmological perturbation theory in a general theory of gravity. Our result is \eqref{eq_gen_turnaround}.
Our derivation makes it clear why the standard formula for the turnaround radius 
(which in fact agrees with eq. 21 of ~\cite{Faraoni2015}) is valid for {\em any theory of gravity obeying the EEP}, once the solution 
for the potential $\Psi$ is known (see eq. \eqref{def_perturbed_FRW_metric} 
and \eqref{reduced_turnaround_equation} below).
In section \ref{sec_BD} we consider a specific theory, the Brans-Dicke theory of gravity, as an example
to demonstrate the use of our formula.  Within the Brans-Dicke theory, we perform the calculation of the turnaround 
radius in two coordinate systems, arriving (as expected) at the same result.
We firstly determine the solution to the field equations around a static spherically mass distribution 
and secondly around a spherical solution in an expanding universe and show that the two solutions are equivalent,
related by a coordinate transformation. Our calculation yields 
$\RTA^{(BD)} \approx \RTA^{(\Lambda CDM)} \left(1 + \frac{1}{3\omega}\right)$ for large Brans-Dicke parameter $\omega$ and hence
it is always larger than $\Lambda$CDM, implying that the Brans-Dicke theory is also consistent with current data.
We conclude finally in section \ref{sec_conclusion}.

Throughout this article we work with mostly positive signature of the metric, $(-,+,+,+)$ and use the greek alphabet for spacetime indices
and latin alphabet for spatial indices. We use units where the speed of light is equal to unity.

\section{The turnaround radius}
\label{sec_turnaround}

\subsection{The turnaround radius in GR with a cosmological constant}
Let us first briefly present the case of GR with a cosmological constant, where the derivation of the turnaround radius is well known.
 This will be useful further below when we generalize the result to arbitrary metric theories of gravity.

In ~\cite{PavlidouTetradisTomaras2014} the turnaround radius for a spherical mass $M$ in the  \LCDM  model was defined in the following way.
 Consider a stationary probe in a Schwarzschild-de Sitter (SdS) spacetime with metric
\begin{equation}
 ds^2_{SdS} = -\left(1 -  \frac{2G_N M}{R}  - \frac{\Lambda}{3} R^2\right) dT^2 + \frac{dR^2}{1 -  \frac{2G_N M}{R} - \frac{\Lambda}{3} R^2 } 
  + R^2 d\Omega
\label{SdS_metric}
\end{equation}
 following a trajectory in spacetime with four-velocity
\begin{equation}
u^\mu = \left( \frac{1}{\sqrt{ 1 -  \frac{2G_N M}{R}  - \frac{\Lambda}{3} R^2 } } , 0, 0 ,0\right).
\label{stationary_vel_static}
\end{equation}
Here $G_N$ is the measured Newtonian gravitational constant.
The maximum turnaround radius is the point along a radial trajectory where the 
four-acceleration $a^\nu = u^\mu \nabla_\mu u^\nu$ of the probe vanishes.
Using the SdS metric \eqref{SdS_metric} yields $ a^1 = \left(1 -  \frac{2G_NM}{R}  - \frac{\Lambda}{3} R^2\right) 
 \left( \frac{G_NM}{R^2}  - \frac{\Lambda}{3} R\right)$ and setting it to zero 
gives the turnaround radius, 
\begin{equation}
\RTA = \left(\frac{3G_NM}{\Lambda}\right)^{1/3}
\label{RTA_LCDM}
\end{equation}
in the case of GR with a cosmological constant.

In a different theory of gravity, the SdS metric  \eqref{SdS_metric} need not be a solution. However, assuming that a static solution exists
of the form 
\begin{equation}
 ds^2 = -f(R) dT^2 + h(R) dR^2 +  R^2 d\Omega
\label{gen_static_metric}
\end{equation}
one can follow the same line of thought to define the turnaround radius by the vanishing of the four-acceleration for a stationary probe.
This leads to the condition 
\begin{equation}
  f'(R) \equiv   \frac{\partial f}{\partial R} \underset{\; \tt{ at }\; R=\RTA \; }{\longrightarrow} 0
\label{gen_static_metric_TA}
\end{equation}
 supplying us with an algebraic equation for $R$, which must be solved in order to obtain the maximum turnaround radius $\RTA$.
The definition \eqref{gen_static_metric_TA} is valid in any theory of gravity which obeys the weak equivalence principle
and can be used to calculate the turnaround radius once the solution $f(R)$ is known. Let us also note that
even if the spacetime is spherically symmetric but not static, the metric may still be brought into a diagonal form, in which case
the condition \eqref{gen_static_metric_TA} still holds, although the resulting turnaround radius will in general be time dependent.

The above definition \eqref{gen_static_metric_TA} of the turnaround radius is not formulated in a covariant language, but can be made so.
In particular, the turnaround radius corresponds to the locus where $u^\mu \nabla_\mu u^\nu = 0$ 
for a \emph{stationary observer} in a spherically symmetric spacetime.  With this definition one 
can calculate the turnaround radius in any coordinate system of choice, although, the definition depends 
on this particular choice of observer.

Our goal is to find a definition of the turnaround radius, suited for cosmology, equivalent to the definition above.
Consider the McVittie metric~\cite{McVittie1933,Gao2004,KaloperKlebanMartin2010}
\begin{equation}
 ds_{McV}^2 = -\left(\frac{1 - \mu}{1+\mu}\right)^2 dt^2  + (1 + \mu)^4  a^2 (dr^2 + r^2 d\Omega)
\label{McVittie_metric}
\end{equation}
where $\mu = \frac{G_NM}{2 a r}$, describing the exterior of a spherical mass in an expanding Universe evolving with
 scale factor $a(t)$. The field equations are
\begin{align}
3H^2 &= 8\pi G_N \rho
\\
   -2 \frac{1+\mu}{1-\mu}\dot{H} - 3H^2  &= 8\pi G_N P
\end{align}
where $H(t) \equiv \dot{a}/a$ is the Hubble parameter, 
$\rho=\rho(t)$ is the energy density and $P=P(t,r)$ the (inhomogeneous) pressure.~\footnote{Having a homogeneous density, yet, inhomogeneous 
pressure seems somewhat unnatural.}
 If $8\pi G_N \rho =  \Lambda$ is a constant
then this spacetime reduces to the Schwarzschild-de Sitter spacetime in a different coordinate system to \eqref{SdS_metric}.
To see this (and remembering always that $H$ is a constant in Schwarzschild-de Sitter) define new coordinates $T(t,r)$ and $R(t,r)$ via
\begin{align}
\label{eq_T_coord}
t &= T - Q(R) \, ,
\\
R &= (1 + \mu)^2 a  r \, ,
\label{eq_R_coord}
\end{align}
with $Q(R)$  the solution to
\begin{equation}
\frac{\partial Q}{\partial R} =  \frac{\sqrt{\frac{\Lambda}{3}} R}{ \left(  1 -\frac{2G_N M}{R}  - \frac{\Lambda}{3} R^2 \right)
  \sqrt{ 1 - \frac{2 G_NM}{R}} } 
\label{Q_prime}
\end{equation}
so that  one recovers \eqref{SdS_metric}.

How does the turnaround condition look-like from the McVittie's point of view?
Since we already know the result in the case of the static Schwarzschild-de Sitter coordinate system, we can simply
transform the conditions leading to that result, to the McVittie coordinate system. In particular, we need to transform the
velocity vector field \eqref{stationary_vel_static} of the stationary observer, 
 into the new system~\footnote{It is easy to show that using a stationary observer in the McVittie coordinate system fails.
 Indeed a stationary observer in one coordinate system is no longer stationary in the other.} and
apply the condition $u^\mu \nabla_\mu u^\nu = 0$.
For this we need the inverse transformation of \eqref{eq_R_coord}, i.e.
\begin{equation}
r(t,R) = \frac{ R - G_NM + \sqrt{ R^2- 2 G_NM R  } }{2a} \, ,
\end{equation}
where we have chosen the positive sign of the square root.~\footnote{The negative sign also works, however, issues arise when 
one considers a perturbative analogue of the McVittie metric as we do further below.}

With the above transformation, the observer's velocity  \eqref{stationary_vel_static} becomes
\begin{align}
u^\mu =  \frac{1+\mu}{\sqrt{ (1 - \mu)^2   - H^2 (1+\mu)^6 a^2 r^2  }} ( 1 , -rH, 0 ,0).
\label{stat_vel_mcvittie_r}
\end{align}
Using the condition $u^\mu \nabla_\mu u^\nu=0$ we find (remember $H$ is constant)
\begin{equation}
 2\mu  = (1+\mu)^6  H^2 a^2 r^2
\end{equation}
which translates to \eqref{RTA_LCDM} using \eqref{eq_R_coord}.

\subsection{New definition of the turnaround radius}
We now present a new definition of the turnaround radius, valid in any theory of gravity obeying the EEP.
In generic alternative theories of gravity that we deal with in this article,
 the Schwarzschild-de Sitter metric will in general not be a solution. Neither will 
some general static spherically symmetric metric  have an equivalent form, which resembles the McVittie metric. However, 
our interest is in cosmology, where a perturbed FRW metric always exists. 
Let us then consider the perturbed version of the McVittie construction of the previous subsection.

In the Newtonian gauge, the perturbed FRW metric takes the form
\begin{equation}
ds^2 = - \left(1 + 2 \Psi\right) dt^2  + a^2 \left( 1 - 2 \Phi\right) \gamma_{ij} dx^i dx^j
\label{def_perturbed_FRW_metric}
\end{equation}
where $\Psi$ and $\Phi$ are the two metric potentials and 
where we have assumed that  $\gamma_{ij}$ is flat, so that $\gamma_{ij} dx^i dx^j = dr^2  + r^2 d\Omega$ in spherical coordinates.

By inspection, when $\mu\ll 1$, the McVittie metric \eqref{McVittie_metric} may be interpreted as a perturbation on 
FRW sourced by a point-mass by identifying $\Psi = \Phi = -2\mu = -\frac{G_NM}{ar}$. We exploit this fact and re-cast the definition 
of the turnaround radius using cosmological perturbation theory.
Starting from  \eqref{stat_vel_mcvittie_r}, we rotate into an arbitrary spatial direction, 
using $r^i = (x,y,z) =  \frac{1}{2} \grad^i r^2$, where $\grad^i = \gamma^{ij} \grad_j$.
 The $3$-vector $r^i$ has components $(-rH,0,0)$ in the original coordinate system 
used in   \eqref{stat_vel_mcvittie_r}. We also use the Friedman equation, $\Lambda = 3 H^2$, 
so that the equivalent version of
\eqref{stat_vel_mcvittie_r} albeit in an arbitrary direction is
\begin{align}
u^\mu &=   \frac{1+\mu}{\sqrt{
 (1-\mu)^2 - (Har)^2(1+\mu)^6 }}  ( 1 , - \frac{1}{2} H \grad^i r^2).
\end{align}
This is the four-velocity of a test particle at rest in a coordinate system which is equivalent to \eqref{def_perturbed_FRW_metric}.
Taking the limit $\mu\ll 1$ and $a H r \ll 1$, corresponding to regions far away from both horizons, leads to
\begin{align}
u^\mu &=  ( 1 -\Psi  + H a \Theta ,   - \frac{1}{a} \grad^i \Theta )
\label{vel_Theta}
\end{align}
 where we have defined the scalar  function
\begin{equation}
\Theta = \frac{1}{2} a H r^2
\label{Theta_sph_sym}
\end{equation}
We have assigned the perturbation orders $\Order(\Psi) \sim \Order(H^2) \sim \Order(\Theta^2) \sim \Order(\dot{\Theta})$, which are
remiscent of the Parametrized Post-Newtonian formalism. Indeed the vector field $\grad^i \Theta$ has all the properties of a 
spatial curl-less velocity field.

We have managed to create a covariant definition of the turnaround radius, which is adapted to cosmology. In 
particular one starts from the observer moving with velocity given by \eqref{vel_Theta} and impose the EEP.
The EEP implies the geodesic equation
\begin{equation}
  u^\mu \nabla_\mu u^\nu =0
\end{equation}
which in turn leads to
\begin{equation}
  \grad_i \left[\dot{\Theta} - \frac{1}{a}  \Psi +  H  \Theta \right] - \frac{1}{a} \grad^j \Theta \grad_j  \grad_i \Theta = 0.
\end{equation}
Since the last term can be written as $ \grad^j\Theta \grad_j \grad_i\Theta = \frac{1}{2} \grad_i |\grad\Theta|^2$,
 we finally get the general \emph{turnaround equation}
\begin{equation}
  \grad_i \left[ a \left(\dot{\Theta} +   H \Theta \right) - \frac{1}{2} |\grad\Theta|^2\right] = \grad_i \Psi
\label{eq_gen_turnaround}
\end{equation}
The above equation is valid in any theory of gravity obeying the EEP. 
 Despite appearances the above equation is fully consistent in perturbation theory (remember the assignment of perturbation orders above).
One should not treat \eqref{eq_gen_turnaround} as a differential equation for $\Theta$. Rather, one should 
assume a specific functional form for $\Theta(\vec{x}^i,t)$ and  then given that functional form, as well as the solution for
$\Psi$ from the field equations of the theory, one should determine the $3$-surface ${\cal F}(x^i) = {\rm const}$ such that the equation holds.
In the case of spherical symmetry $\Theta$ is given by \eqref{Theta_sph_sym}, however,  \eqref{eq_gen_turnaround} may be used
as a starting point for generalizing the turnaround radius calculation into a turnaround surface when the shape of the bound object
is non-spherical. One possibility would be to consider a non-spherical function $\Theta(t,\vec{x})$
 corresponding to some non-spherical surface.

Let us now return to our spherically-symmetric ansatz, i.e. $\Theta = \frac{1}{2}  a H r^2$. In this case we have that 
$|\grad\Theta|^2 =  a^2 H^2 r^2 = 2  a H \Theta$, hence the LHS of \eqref{eq_gen_turnaround} leads to
\begin{equation}
    a \frac{\partial \dot{\Theta}}{\partial r} =  a^2 [ H^2  +  \dot{H} ] r 
\label{reduced_turnaround_equation_LHS}
\end{equation}
and the turnaround equation simplifies to
\begin{equation}
 a^2 [ H^2  +  \dot{H} ] r = \frac{\partial \Psi}{\partial r}.
\label{reduced_turnaround_equation}
\end{equation}
The above equation which we name the \emph{reduced turnaround equation} (due to spherical symmetry) 
can then be used to calculate the turnaround radius $\RTA = a r$ given a Hubble parameter $H(t)$ and the solution
to the potential $\Psi$, both of which are specified in a given theory, including a theory beyond GR.

From \eqref{reduced_turnaround_equation} a quick calculation gives the turnaround radius 
for the case of a  cosmological constant as dark energy
 and for the case of a dark energy fluid with equation of state parameter $w$  both within the GR framework.
In both models the solution to the potential is $\Psi  = -\frac{G_NM}{ar}$ ~\cite{PavlidouTetradisTomaras2014}. 
 What is different between the two models is the Hubble parameter.
In the first case it is a constant given by $ H = \sqrt{\Lambda/3}$ so that \eqref{reduced_turnaround_equation} leads to \eqref{RTA_LCDM},
while in the second case it is given by $a H = H_0 a^{-(1+3w)/2}$  so that (since for dark energy $1+3w<0$)
\begin{equation}
 \RTA  =  \left[- \frac{2G_NM}{(1+3w)H^2} \right]^{1/3} = \left[- \frac{2G_NM}{(1+3w)H_0^2} \right]^{1/3} a^{1+w} 
\end{equation}
We observe that when $w\ne -1$ the maximum turnaround radius is time-dependent. In the limit $w\rightarrow -1$, i.e. $\Lambda$CDM,
 the maximum  turnaround radius agrees with the time-independent $\Lambda$CDM formula \eqref{RTA_LCDM}.

\section{The turnaround radius of Brans-Dicke theory of gravity}
\label{sec_BD}
The Brans-Dicke theory~\cite{BransDicke1961} can be thought of as a prototype  alternative theory of gravity.
Its action in the presence of a cosmological constant is given by
\begin{eqnarray}
S= \frac{1}{16\pi G} \int \sqrt{-g} d^4 x \left[\phi R  - 2 \Lambda -\frac{\omega}{\phi} (\nabla\phi)^2 \right] + S_M,
\label{BD_action}
\end{eqnarray}
where the scalar $\phi$ is the Brans-Dicke field, the constant $\omega$ is the Brans-Dicke parameter 
and $S_M$  is the  collective action for all matter fields present, which depends on the metric $g_{\mu\nu}$ but not on the scalar field.
The shift of conceptual paradigm from GR in this theory is certainly the scalar field $\phi$, 
whose non-minimal coupling with the Ricci scalar indicates a spacetime
dependent gravitational coupling. In the limit $\omega\rightarrow\infty$ 
the scalar field $\phi$ must be a constant $\phi\rightarrow \phi_0$ in which case GR is recovered.

Solar system data severely constrain  $\omega \gtrsim 40 000$~\cite{BertottiIessTortora2003,Will2014}, 
thereby making it practically indistinguishable from General Relativity in our local neighbourhood. 
However, any test of gravity should be accompanied by a specification of the curvature and potential 
regime it is performed in~\cite{BakerPsaltisSkordis2015}. In this sense cosmological constraints on Brans-Dicke theory 
should be treated independently from solar system tests as they lie in different regions of the gravitational parameter space.

Let us exemplify. As shown in ~\cite{AvilezSkordis2013},
the Brans-Dicke theory arises as a specific limit of Horndeski theory~\cite{Horndeski1974,DeffayetGaoSteer2011},
the most general Lorentz-invariant scalar-tensor theory, having second order field equations in four dimensions. 
The Horndeski theory offers the possibility of realizing screening mechanisms
such as the Vainshtein~\cite{Vainshtein1972}, the chameleon~\cite{KhouryWeltman2003} and the symmetron~\cite{HinterbichlerKhoury2010}
 mechanisms. 
These mechanisms restore GR around the high-curvature/high-density environments of astrophysical bodies, such as the sun. 
Hence, it is possible that certain subsets of Horndeski theory which realize these mechanisms tend to Brans-Dicke theory
in the low curvature environment of the cosmological regime but acquire corrections which send it back to GR in regions
of high curvature.
As such, cosmological constraints on Brans-Dicke theory give different information than solar system tests.
In \cite{AvilezSkordis2013} the lower bound $\omega > 890$ at the $99\%$ confidence level was placed (see 
also \cite{LiEtAl2015,BallardiniEtAl2016}),
using the latest Cosmic Microwave Background  data from Planck. Future photometric and spectroscopic cosmological surveys
are expected to increase this by a factor of $20 - 30$~\cite{AlonsoEtAl2016,AvilezSkordisSong2016}, 
making cosmological tests comparable to solar system tests.

In~\cite{BhattacharyaEtAl2015}, the no hair theorems for the Brans-Dicke theory with $\Lambda>0$ for stationary axisymmetric black holes 
and stars were discussed. It was shown there that no matter how large the Brans-Dicke parameter $\omega$ is, 
unless it is {\it infinite} (i.e., the theory coincides exactly with the  General Relativity), there can exist no 
regular such solutions if asymptotic de Sitter boundary condition is imposed. 
The Brans-Dicke theory has also been  investigated  in the context of galactic dark matter in~\cite{DeyBhattacharyaSarkar2014}. 

In order to pave the way for the calculation of the turnaround radius we construct solutions in Brans-Dicke theory with a cosmological constant.
We consider two types of solutions, i.e. static spherically symmetric solutions and cosmological solutions, in order to
apply both formulae \eqref{gen_static_metric_TA} and \eqref{reduced_turnaround_equation} for the determination of the turnaround radius.

\subsection{Stationary spherically symmetric point-mass solutions}
\label{sec_BD_SSS}
Adopting a static spherically symmetric ansatz as in \eqref{gen_static_metric} and in addition that $\phi = \phi(R)$,
the field equations are 
\begin{subequations}
\begin{align}
&
  \frac{1}{R} \left( \frac{h'}{h}  +\frac{h - 1 }{R}   \right)  
- \frac{1}{2}\omega \left( \frac{\phi'}{\phi} \right)^2
+    \frac{f'}{2 f}  \frac{\phi'}{\phi}
 = \frac{ 8\pi G  h }{\phi}  \left[ \rho +   \frac{-\rho + 3P}{2\omega+3}   \right]
\label{eq_E_sp_00}
\\
&
\frac{1-h}{R^2}  +   \frac{1}{R}  \frac{f'}{f}  
+  \frac{2}{R} \frac{ \phi' }{\phi}
+  \frac{f'}{2f} \frac{\phi' }{\phi}
- \frac{1}{2}\omega \left( \frac{ \phi' }{\phi}  \right)^2
 =  \frac{8\pi G P}{\phi}  h 
\label{eq_E_sp_11}
\\
&
 \frac{f''}{2f}  - \frac{(f')^2}{2f^2}  
   + \left( \frac{f'}{2f}   - \frac{h'}{2h}   \right) \left(\frac{f'}{2f}   +  \frac{1}{R}   \right)
+  \frac{1}{2} \omega \left(\frac{\phi'}{\phi}\right)^2
 - \frac{1}{R}  \frac{\phi'}{\phi} 
 = \frac{8\pi G h}{\phi} \left[P  +  \frac{\rho-3P}{2\omega+3}   \right]
\end{align}
and
\begin{equation}
  \left[ \frac{\sqrt{f}}{\sqrt{h}} R^2    \phi' \right]'    =  \frac{8\pi G \left(\rho - 3 P\right)}{2\omega+3}  \sqrt{fh} R^2 
\label{scalar_sp_sym}
\end{equation}
\label{sp_sym}
\end{subequations}
where $\rho$ and $P$ are  the total density and pressure of matter respectively, including the cosmological constant.
Consistency requires that the matter velocity has components
$u^\mu = (\frac{1}{\sqrt{f}} ,0 ,0 ,0)$. 
In the Einstein equations above, we have used the scalar equation \eqref{scalar_sp_sym} to
eliminate the $\square \phi$ terms.

A complete analytic solution of \eqref{eq_E_sp_00}-\eqref{scalar_sp_sym} is impossible. Indeed, as we discussed above,
 it has been shown that the Brans-Dicke theory with a cosmological constant does not admit stationary 
and spherically symmetric solutions, which are exterior solutions to a compact object 
and which have a cosmological horizon where the Brans-Dicke field is regular~\cite{BhattacharyaEtAl2015}. 
Clearly then, any spherically symmetric solution in this theory (in the presence of $\Lambda$) must be necessarily time-dependent. 
However, we expect this time-dependence to become more and more manifest only when we approach the 
cosmological horizon. As the turnaround radius is on much smaller scales, we take a different approach: perturbation theory.

Physical systems of interest are those where the Schwarzschild horizon $R_s$,
the turnaround radius $\RTA$ and the de Sitter horizon $R_h$ are widely separated. To be more precise, 
in standard GR we have $R_s/\RTA = 2G_N M  (\frac{\Lambda}{3G_NM})^{1/3} \lesssim 10^{-8}  -  10^{-4}$  for the most massive 
galaxy clusters in the range $M \sim 10^{11} - 10^{17} M_\odot$ while 
$\RTA/R_h =  (\frac{3G_NM}{\Lambda})^{1/3} \sqrt{\Lambda}/\sqrt{3} \lesssim  10^{-4} - 10^{-2}$. It thus seems like a good first 
approximation that $2G_NM/R \ll 1$ and $ \Lambda R^2/3 \ll 1$, so that the Scharzschild-de Sitter spacetime may be considered as a perturbation around
Minkowski for the scales of interest.\footnote{
One may instead perturb around a de Sitter, or even, a Schwarzschild-de Sitter spacetime. However, this introduces tremendous complication
in solving the scalar equation and in the end, the Minkowski space approximation used  here, where $2G_NM/R \ll 1$ and $ \Lambda R^2/3 \ll 1$,
is recovered.}

We expand our variables as
\begin{align}
 f &= 1 + U
\\
 h &= 1 + V
\\
 \phi &= \phib_0( 1 + \varphi )
\end{align}
so that $U$, $V$ and $\varphi$ are small compared to unity and $\phib_0$ is a background value for $\phi$.
We consider a point-mass source in a spacetime filled with a cosmological constant  so that the energy-density  and pressure entering
 \eqref{eq_E_sp_00}-\eqref{scalar_sp_sym}  take the form
\begin{align}
8\pi G \rho &= \frac{2GM}{R^2} \delta(R) + \Lambda
\\
8\pi G P &=  - \Lambda
\end{align}
Consistently with our approximation both the point mass and $\Lambda$ are treated as small perturbations.
We start from  \eqref{eq_E_sp_00}, linearize and then integrate to get
\begin{equation}
  V
 = \frac{1}{\phib_0(2\omega+3) }  \left[ 
  2(\omega+1) \frac{2GM }{R} 
+  \frac{2\omega - 1 }{3}  \Lambda R^2
  \right].
\end{equation}
The above solution is then used in the linearized version of \eqref{scalar_sp_sym}, which when integrated gives
\begin{equation}
    \varphi     =  -\frac{1}{\phib_0(2\omega+3)}  \left[ \frac{2}{3}\Lambda R^2 -  \frac{2GM}{R} \right].
\label{varphi_static}
\end{equation}
Finally, the expressions for $V$ and $\varphi$ are used in the linearized version of  \eqref{eq_E_sp_11}, leading after integration to
\begin{equation}
  U  =  
-  \frac{1}{\phib_0(2\omega+3) }  \left[ 
   2(\omega+2) \frac{2GM }{R} 
+  \frac{2\omega+1}{3} \Lambda  R^2  
  \right]
\end{equation}
so that the metric is
\begin{align}
 ds^2 &= -\left[1  
-  \frac{2(\omega+2)}{2\omega+3}  \frac{2GM }{\phib_0 R} -  \frac{2\omega+1}{2\omega+3} \frac{\Lambda}{3\phib_0}  R^2  \right] dT^2
\nonumber
\\
& 
 + \left[1 
+  \frac{2(\omega+1)}{2\omega+3}  \frac{2GM }{\phib_0 R} 
+  \frac{2\omega-1}{2\omega+3}     \frac{\Lambda}{3\phib_0}  R^2 \right] dR^2
+  R^2 d\Omega
\label{static_BD_metric_epsilon}
\end{align}

\subsection{Cosmological solutions with a point-mass source}
\label{sec_BD_cosmo}
Let us now construct cosmological solutions for the metric and the Brans-Dicke field with a point-mass source. 
By ``cosmological'' we mean that in the limit $M\rightarrow 0$, the metric becomes the Friedman-Robertson-Walker metric and so these
solutions are the analogue of the McVittie solution in the case of GR.
We construct our solution by first considering a  background FRW solution and then adding the perturbation due to the mass 
(see also \cite{Eingorn2015} for  cosmological perturbation theory equations with an array of point masses).

\subsubsection{FRW solutions}
\label{sec_BD_FRW}
The FRW metric is
\begin{equation}
ds^2 =   - dt^2 +  a^2 \gamma^{(\kappa)}_{ij} dx^i dx^j,
\label{def_FRW_metric}
\end{equation}
where $a(t)$ is the scale factor of cosmic time $t$, $\gamma^{(\kappa)}_{ij}$ is the 3-metric (used to raise and lower three-dimensional indices)
of constant spatial curvature $\kappa$.

The Friedman equation in Brans-Dicke theory takes the form
\begin{equation}
3\left( H + \frac{1}{2} \frac{\dot{\phib}}{\phib} \right)^2
+   \frac{3\kappa}{a^2} = \frac{8\pi G}{\phib} \rhob
 +  \frac{2\omega+3}{4} \left(\frac{\dot{\phib}}{\phib}\right)^2 
\label{friedman_1}
\end{equation}
where $\rhob$ is the background energy-density of matter (including the cosmological constant), $H= \dot{a}/a$ 
 is the time-dependent Hubble parameter and $\phib$ is the homogeneous part of the scalar field adopted to the 
FRW symmetries.  The scalar evolves according to
\begin{equation}
\ddot{\phib} + 3 H \dot{\phib} = \frac{8\pi G}{2\omega+3} ( \rhob - 3 \Pb) 
\label{scalar_FRW}
\end{equation}
where $\Pb$ is the background pressure of matter (including the cosmological constant). 

It is straightforward to verify that an exact analytical solution is, in general, impossible, even if
$8\pi G \rhob = \Lambda $ is a constant. Indeed, it can be shown that the de Sitter spacetime in no
longer an exact solution of the field equations as it is in GR.
\footnote{It may be shown that an exact solution exists for $8\pi G \rho = \Lambda = {\rm const}$
with $a = \left(t/t_0\right)^{\frac{2\omega+1}{2}}$ and $\phi = 4 \Lambda t^2 /(2\omega+3)/(6\omega+5)$. However, this requires that
 initially both the scalar and its first derivative vanish, i.e. $\phib_0 = \dot{\phib}^{(in)} = 0$,
 and therefore this is a spurious solution of no physical significance and must be discarded. 
}. This is equivalent to the non-existence of static-spherically symmetric solutions in the presence of a
cosmological constant~\cite{BhattacharyaEtAl2015}, as we have discussed in the previous subsection.
Hence, we proceed using perturbation theory.  Both the Friedman equation \eqref{friedman_1} and the scalar equation \eqref{scalar_FRW} suggest
that the small parameter to use is 
\begin{equation}
\epsilon = \frac{1}{2\omega+3}
\end{equation}

We are interested in the case of a flat universe filled with cosmological constant so that $3 H_0^2 \phib_0 = 8\pi G \rhob = \Lambda = -8\pi G \Pb$.
We construct the perturbative solution as a power series in $\epsilon$ which yields
\begin{align}
\phib &= \phib_0\left[ 1  + 4  \epsilon  \ln a + \ldots \right]  = \phib_0\left[  1  + 4 \epsilon H_0 t + \ldots \right]
\label{phi_FRW_lna_Lambda}
\\
H &= H_0  \left[ 1- \frac{4}{3}  \epsilon -  2  \epsilon \ln a + \ldots \right] 
  = H_0 \left[ 1- \frac{4}{3}  \epsilon -  2  \epsilon H_0 t + \ldots \right]
\label{H_FRW_lna_Lambda}
\end{align}
as can be verified by direct substitution. A more formal derivation which is valid for a generic
matter field and curvature can be found in the appendix.
The dependence of the scale factor on time $t$ is found by integrating \eqref{H_FRW_lna_Lambda} so that
 to $\Order(\epsilon)$ we find
\begin{equation}
a = \abar \left[ 1   - \frac{4}{3} \epsilon H_0 t - \epsilon   (H_0 t)^2 + \ldots \right]
\label{a_FRW_t}
\end{equation}
where $\abar = e^{H_0 t}$. The solutions found above are of course only valid close to $\ln a \sim 1$, i.e.  for all times $t$ 
such that $\epsilon H_0 t \ll 1$.

\subsubsection{Perturbed FRW solutions}
\label{sec_BD_perturbed_FRW}
Including the point-mass in our system inevitably introduces spatial dependence in the solutions.
Assuming that the point-mass is not too massive as to overclose the universe, we may treat its contribution as a perturbation on top
of the FRW solution we have constructed. This requires perturbing the FRW metric 
to linear order as in \eqref{def_perturbed_FRW_metric} by adopting the Newtonian gauge. Likewise we perturb the scalar field
\begin{equation}
\phi = \phib ( 1 + \varphi )
\end{equation}
where $\phib$ is the background value and $\phib \varphi$ the perturbation.

Before proceeding into solving the system, caution is warranted.
Our background solution was arbitrarily close to de Sitter. We may then re-interpret the background FRW solution
as being exact de Sitter plus small time-dependent perturbations.
In other words  set $a = \abar(1 + \delta_a)$ where from \eqref{a_FRW_t} 
we get $\delta_a  = - \epsilon\left[ \frac{4}{3}  H_0 t + (H_0 t)^2\right]$.  This also implies $H = H_0 + \dot{\delta}_a$,
which may be checked for consistency with \eqref{H_FRW_lna_Lambda}.
 Then we may define a new potential as $a^2 (1 - 2\Phi) = \abar^2 (1 - 2 \Phit)$ so that $\Phit =  \Phi -  \delta_a$. The background field equations
can only be satisfied under this transformation, if and only if a further transformation is also implemented:
by observing that $\phib = \phib_0(1 + \delta_\phi)$ with $\delta_\phi =  4\epsilon H_0  t$  from \eqref{phi_FRW_lna_Lambda},
 we may transform $\delta_\phi$ away via $\phi = \phib (1 + \varphi ) = \phib_0(1 + \varphit)$ so that $\varphit = \varphi +  \delta_\phi$.

Consistency of this line of thought requires that $\Order(\Phi) \sim \Order(\Phit) \sim \Order(\delta_a) \sim \Order(\delta_\phi)$ so
 that when considering linearized perturbations we ignore terms like $\Phi \delta_a$ or $\delta_a^2$, etc. This means that in the
 perturbation equations we may replace  $a\rightarrow \abar$, $H\rightarrow H_0$ and $\dot{\phib} \rightarrow 0$
 resulting in great simplification. A further consistency requirement is that since after the transformation the background scalar field is
constant, the scalar field equation must be treated entirely perturbatively. 
With these considerations and letting $ \grad_i $ to be the covariant derivative of $\gamma_{ij}$,
the perturbed Einstein equations sourced by matter with density perturbation $\delta \rho = M \delta^{(3)}\left( a \vec{r}\right) $ 
are as follows.  Using the identity $\delta^{(3)}\left( a \vec{r}\right) = \delta(r)/ (4\pi a^3 r^2)$, 
the $0-0$  perturbed Einstein equation is
\begin{align}
&
-6  H_0 \left(\dot{\Phit} + H_0   \Psi \right)
+ 3 H_0 \left(\dot{\varphit} + H_0 \varphit \right)
 + \frac{2}{\abar^2} \grad^2  \left(\Phit - \frac{1}{2} \varphit  \right)
 =
\frac{2 G M }{\phib_0 \abar^3 r^2}  \delta(r)
\label{G_00_pert_FRW}
\end{align}
and the $0-i$-Einstein equation is
\begin{equation}
2\grad_i \left( \dot{\Phit} + H_0  \Psi  \right) = \grad_i \left( \dot{\varphit} -  H_0  \varphit \right).
\label{G_0i_pert_FRW}
\end{equation}
We combine \eqref{G_00_pert_FRW} and \eqref{G_0i_pert_FRW}, assume the quasistatic limit where $H_0^2 \varphit \ll \grad^2 \varphit$ 
and integrate to get
\begin{align}
&
  \Phit - \frac{1}{2} \varphit  = - \frac{ G M }{\phib_0 \Rb} 
\end{align}
where we have defined
\begin{equation}
 \Rb  = \abar r.
\end{equation}
The perturbed scalar field equation is
\begin{equation}
 \ddot{\varphit} 
+  3 H_0  \dot{\varphit} 
- \frac{1}{\abar^2} \grad^2 \varphit
 = 
\frac{\epsilon}{\phib_0} \left[
 4\Lambda  
+  \frac{2 G M }{ \abar^3 r^2}\delta(r) 
+ 8\Lambda    \Psi
\right]
\label{scalar_pert_FRW}
\end{equation}
and after assuming the quasistatic limit and integrating gives
\begin{equation}
   \varphit =  2 \epsilon \left[\frac{G M}{\phib_0 \Rb}  -   H_0^2  \Rb^2 \right]
\label{varphit_cosmo}
\end{equation}
Hence, $\Phit =   - \frac{ G M (1 - \epsilon) }{\phib_0 \Rb} -\epsilon   H_0^2  \Rb^2$
while $\Psi  = - \frac{ G M (1 + \epsilon) }{\phib_0 \Rb} +\epsilon   H_0^2  \Rb^2 $ after using the 
 traceless-$ij$-Einstein equation $D_{ij} \left(\Phi - \Psi -   \varphi  \right) = 0$ and ignoring the kernel which results to 
pure gauge-solutions.

Therefore, the metric to $\Order(\epsilon)$ is
\begin{equation}
ds^2 =  
  - \left[1-\frac{2GM}{\phib_0 \Rb}(1+  \epsilon ) + 2 \epsilon   H_0^2  \Rb^2  \right] dt^2 
+ \abar^2 \left[1+ \frac{2GM}{\phib_0 \Rb }(1-\epsilon  )  + 2 \epsilon   H_0^2  \Rb^2 \right]\gamma_{ij} dx^i dx^j 
\label{metrict_cosmo}
\end{equation}
Setting $\epsilon\rightarrow 0$ recovers the perturbed McVittie metric as expected, i.e. it recovers \eqref{McVittie_metric} in the limit
$\mu\ll 1$.

\subsection{The turnaround radius in Brans-Dicke theory}
Having found the two types of solutions let us return to our original goal: the turnaround radius.
A quick calculation using \eqref{gen_static_metric_TA} along with 
the static spherically symmetric solution \eqref{static_BD_metric_epsilon} yields
\begin{align}
  \RTA^3 = \frac{3GM}{ \Lambda}\frac{2\omega+4}{2\omega+1}  
\end{align}
and taking the large $\omega$ (small $\epsilon$) limit
\begin{equation}
\RTA \approx  \left( \frac{ 3 G M}{\Lambda } \right)^{1/3}  (1+ \epsilon) \approx  \left( \frac{ 3 G M}{\Lambda } \right)^{1/3}  
 \left(1 + \frac{1}{2\omega}\right) 
\label{RTA_BD_raw}
\end{equation}
to $\Order(\epsilon) \sim \Order(1/\omega)$.

Similarly, another quick calculation using \eqref{reduced_turnaround_equation}
 along with $H = H_0$ and the cosmological solution \eqref{metrict_cosmo} yields once again \eqref{RTA_BD_raw}.
This should not come as a surprise. After all the two solutions \eqref{static_BD_metric_epsilon} 
and  \eqref{metrict_cosmo} are in fact one and the same, after a coordinate transformation.
This may be checked using the general form of such coordinate 
transformations between a static spherically symmetric space time and a perturbed FRW spacetime~\cite{Skordis2016}. 

Note that we may also transform the cosmological solution 
back to the original FRW background  given by \eqref{phi_FRW_lna_Lambda}, \eqref{H_FRW_lna_Lambda} 
and \eqref{a_FRW_t}. In that case, the potential $\Phi$ acquires a pure time-dependence, which is in turn eliminated by a gauge-transformation.
This introduces a time-dependence into $\Psi$ and in order to use \eqref{reduced_turnaround_equation}
we must determine the canonical form of $\Psi$ as in \cite{Skordis2016}. This is found to be 
$\Psi = - \frac{ G M (1 + \epsilon) }{\phib_0 R} -  \epsilon H_0^2 ( \frac{4}{3} +  2H_0 t ) R^2 $ so that \eqref{reduced_turnaround_equation} along with \eqref{H_FRW_lna_Lambda}  gives back \eqref{RTA_BD_raw}.

In \eqref{RTA_BD_raw} we have found the turnaround radius in terms of the bare parameters of the theory, $G$ and $\Lambda$. However,
as is well known, the bare $G$ in the Brans-Dicke action is not the  actual measured Newtonian gravitational constant $G_N$.
Indeed, the latter is defined  as~\cite{Weinberg1972,Will1981,Will2014}
\begin{equation}
 G_N =  \frac{2(\omega+2)}{\phib_0(2\omega+3) }    G 
\approx  \frac{1 + \epsilon}{\phib_0} G ,
\end{equation}
so that  $g_{00} \approx -1 + 2G_N M / R$ as $R\rightarrow 0$.  Hence, $3G M / \Lambda = (1-\epsilon) G_N M / H_0^2$.
Furthermore, we should consider  how we measure the cosmological constant. The Friedman equation (under the assumption
 that $\phi \approx {\rm const}$) is $3H^2 \approx  \Lambda /\phib_0 + 8\pi G_N (1 - \epsilon) \rho_{\rm matter} $.  
Hence, using cosmological observations one would measure $\Lambda_{\rm eff} = \Lambda /\phib_0$ rather than the bare $\Lambda$ and 
we call this the effective cosmological constant.
With these considerations the expression \eqref{RTA_BD_raw} should be adjusted accordingly to
\begin{equation}
\RTA \approx  \left( \frac{ 3 G_N M}{\Lambda_{\rm eff} } \right)^{1/3}  \left(1+ \frac{2}{3} \epsilon\right) \approx  
\left( \frac{ 3 G_N M}{\Lambda_{\rm eff} } \right)^{1/3}  
 \left(1 + \frac{1}{3\omega}\right)
\label{RTA_BD_res}
\end{equation}
which is our final result.

\section{Conclusions}
\label{sec_conclusion}
In this article we have calculated the effect of generic alternative theories of gravity obeying the Einstein Equivalence Principle
 on the maximum size of  large scale cosmic structures.  The maximum size of a structure is 
given by the maximum turnaround radius $\RTA$ -- the point where the attraction due to the central mass gets balanced 
with the repulsion due to the dark energy, beyond which no compact mass distribution is possible. Thus any model predicting 
a maximum size of a structure with a given mass smaller than its actual observed size, gets ruled out on the basis of the 
stability of the structure. Conversely, if a given theory predicts a maximum size larger than the actual or observed size, 
the theory certainly persists.  The theoretical prediction of $\Lambda{\rm CDM}$ on $\RTA$ was shown to be 
absolutely consistent with the observed astrophysical data~\cite{PavlidouTomaras2013,PavlidouTetradisTomaras2014}, and it is only 
about $10\%$ larger than the observed ones for large scale structures with $M\geq 10^{13}M_{\odot}$ which are yet 
to virialize and much larger for masses below that~\cite{TanoglidisPavlidouTomaras2014}. Thus, it is 
clear that in order to have a meaningful phenomenology with the maximum turnaround radius to constrain various models, we must consider 
large scale objects with $M\geq 10^{13}M_{\odot}$. In particular, such consideration completely rules out 
dark energy models with equation of state parameter $w<-2$~\cite{PavlidouTetradisTomaras2014}.

We have introduced a new definition of the maximum turnaround radius, given by the turnaround equation \eqref{eq_gen_turnaround},
valid in any theory of gravity obeying the EEP and for any non-spherical bound object.
We have further adopted  \eqref{eq_gen_turnaround} under the simplified assumptions of a spherically symmetric setup
and a time-dependent cosmological setup with spherically symmetric perturbations arriving at the same conclusions. 
 In both cases we deal with spherical symmetry.  As we discussed above, since the large scale structures we should 
apply the turnaround calculation to are yet to virialize,  spherical symmetry seems to be a very good approximation for our current purpose.
 The members of such a structure would redistribute their kinetic energy in order to reach virialization and the structure would get 
smaller in size. Thus,  non-sphericity would eventually be created, but at a later time.  In particular, it was argued 
in~\cite{PavlidouTomaras2013} that even the maximum departure from non-sphericity is not very large for most of those structures, 
except that of the Corona-Borialis supercluster -- which may not be a single structure at all. Nevertheless, it is quite 
instructive and interesting to extend the current formalism to include non-sphericity as well. One possibility
is to start from the general turnaround equation~\eqref{eq_gen_turnaround} and consider a non-spherical function $\Theta(t,\vec{x})$,
possibly corresponding to some non-spherical surface. Another possible way to do this without adhering to perturbation theory, would be to 
consider an axisymmetric generalization of the McVittie solution we investigated by putting in a rotation and also to consider the 
Sheth-Tormen statistical mass function instead of the Press-Schechter  statistical mass function  (see e.g.~\cite{Padmanabhan1993}) 
in the analysis of~\cite{TanoglidisPavlidouTomaras2014}.

The most important point we have demonstrated is that the turnaround radii
predicted by both spherically symmetric and cosmological spacetimes are the same -- 
establishing it as a purely geometric, coordinate invariant quantity. Such equality was earlier 
established for \LCDM in~\cite{PavlidouTomaras2013,PavlidouTetradisTomaras2014}.
As an application, we used the formalism in the context of the Brans-Dicke theory with a positive cosmological constant. 
Owing to the severe constraint of the Brans-Dicke parameter from the solar system data, $\omega \gtrsim 40 000$~\cite{Will2014}, 
we used a perturbative expansion in the Brans-Dicke parameter in terms of $\epsilon=1/(2\omega +3)$ and  showed that 
 the maximum turnaround radius is always larger than that of the $\Lambda{\rm CDM}$,  Eq.~(\ref{RTA_BD_res})
since our formula is only valid for $\omega \gg 1$.
The increment of $\RTA$ from the $\Lambda{\rm CDM}$ is apparent from Eq.~(\ref{RTA_BD_res}) -- depicting the 
increment of the term $G_N M$ for a finite and positive $\omega$, keeping $\Lambda$ fixed. The physical meaning 
behind this is related to the fact that since the gravitational attraction in Brans-Dicke is increased compared to
 GR (due to the additional scalar mediating gravity), we should move 
further radial distance away than $\Lambda{\rm CDM}$ in order to get it balanced by the repulsion of the dark energy 
whose value is being fixed.   
In other words, the maximum size of a structure with given mass should be regarded as the maximum length 
scale up to which it can hold itself against the repulsion due to the ambient dark energy. If we specify the latter,
 certainly $\RTA$ would increase with increasing mass or gravitational coupling. 

Another important point to note here that we have used the definition of the mass and the cosmological constant as that of the 
General Relativity in~Eq.~(\ref{RTA_BD_res}). Certainly, this should not be the case in general and such parameters 
should be defined within the framework of the theory itself. However, as long as we are doing perturbation theory over $\Lambda{\rm CDM}$, 
such notion seems practically reasonable. Similar considerations within the Brans-Dicke theory in the context of the  
Parameterized Post-Newtonian formalism can be found in~\cite{Weinberg1972,Will1981}.  In any case, our result 
shows that the Brans-Dicke theory is perfectly consistent with the mass versus observed maximum sizes and hence the stability of structures. 

It would be highly interesting to go beyond the first order perturbation theory considered here, in order to further 
investigate the stability issues. We hope to return to this in a future work.

\acknowledgments 
We thank D.~Tanoglidis for useful conversations.
T. Tomaras was supported by the ``ARISTEIA II" Action of the Operational Program 
``Education and Lifelong Learning" and was co-funded
by the European Social Fund (ESF) and Greek National Resources. S. Bhattacharya 
was also partly supported by this grant, while he was a post doctoral
researcher at ITCP and Department of Physics of the University of Crete, Heraklion, Greece.
K. Dialektopoulos acknowledges support from the COST Action CA15117 (CANTATA) 
and INFN Sez. di Napoli (Iniziative Specifiche QGSKY and TEONGRAV).
A.~E. Romano was supported by the Dedicacion exclusica and Sostenibilidad programs at UDEA, 
the UDEA CODI project IN10219CE and 2015-4044, and Colciencias mobility project COSMOLOGY AFTER BICEP.
The research leading to these results has received funding from the European Research Council under the European
 Union's Seventh Framework Programme (FP7/2007-2013) / ERC Grant Agreement n. 617656 ``Theories
 and Models of the Dark Sector: Dark Matter, Dark Energy and Gravity.''

\appendix

\section{Perturbative solution of the background FRW in Brans-Dicke theory}
In this appendix we give a formal derivation of the perturbative background FRW solution presented in section \ref{sec_BD_FRW}.
We give the derivation for a general matter source in the presence of curvature and specialize at the end to a constant-$w$ component
in a flat universe.

We eliminate the $t$-dependence in the background field equations by changing variables from $t$ to $\ln a$ 
so that the Friedman equation \eqref{friedman_1} becomes
\begin{equation}
H^2
= \frac{\frac{8\pi G}{\phib} \rhob -  \frac{3\kappa}{a^2} }{  3 \left( 1 + \frac{1}{2} \frac{d\ln \phib}{d\ln a} \right)^2
 -  \frac{1}{4} \frac{1}{\epsilon}  \left( \frac{d\ln \phib}{d\ln a} \right)^2 }
\label{friedman_2}
\end{equation}
 while the scalar equation \eqref{scalar_FRW} can be formally integrated to 
\begin{equation}
 \phib  = \phib_0 +   \epsilon \; 8\pi G \int d\ln a \frac{1}{a^3} \frac{1}{H}  \int  d\ln a ( \rhob - 3 \Pb)  \frac{a^3}{H} 
\label{eq_phi_formal}
\end{equation}
We have set the initial condition $\dot{\phib}^{(in)} $ to zero as it leads to a decaying solution. 

The calculation now proceeds by expanding the fields as
\begin{align}
 \phib &= \phib_0\left(1 + \sum_{n=1}^\infty \phib_n \epsilon^n \right)
\\
 H &= \bar{H}  \left( 1 + \sum_{n=1}^\infty  h_n \epsilon^n \right)
\end{align}
where $\bar{H}$ is the time-dependent Hubble parameter in the limit $\epsilon \rightarrow 0$ (not to be confused with the Hubble constant $H_0$)
and is given by $3\bar{H}^2  = \frac{8\pi G}{\phib_0} \rhob -  \frac{3\kappa}{a^2}   $.

Let us define the operator $S[A,B]$ acting on functions $A$ and $B$ by
\begin{equation}
S[A,B] = \frac{8\pi G}{\phib_0} \int d\ln a \frac{1}{a^3} \frac{1}{\bar{H}} A \int  d\ln a ( \rhob - 3 \Pb)  \frac{a^3}{\bar{H}} B
.
\label{S_op}
\end{equation}
This operator is then used to construct the perturbed variables $\phib_n$  from the  scalar integral \eqref{eq_phi_formal}. 
The first three expansion coefficients are
\begin{subequations}
\begin{align}
\phib_1 &= S[1,1]
\label{phi_FRW_pert_1}
\\
\phib_2 &= - S[1,h_1] - S[h_1,1]
\label{phi_FRW_pert_2}
\\
\phib_3 &= S[1, h_1^2 - h_2 ] + S[h_1,h_1] + S[ h_1^2 - h_2 , 1] 
\label{phi_FRW_pert_3}
\\
\ldots
\nonumber
\end{align}
\label{phi_FRW_pert}
\end{subequations}
The Friedman equation \eqref{friedman_2} is also perturbed to give
\begin{subequations}
\begin{align}
  h_1 &= - \frac{1}{2}  \chib_1 \left( 1 - \frac{1}{12}  \chib_1 \right) - (1 - \Omega_K) \frac{1}{2}     \phib_1 
\\
  h_2 &=
 - \frac{1}{2}  \left( \chib_2 -  \phib_1 \chib_1 + \frac{1}{4}  \chib_1^2 \right)
+ \frac{1}{12}  \chib_1 \left( \chib_2 -  \phib_1 \chib_1 \right)
 + \frac{3}{8}  \chib_1^2 \left( 1 - \frac{1}{12}  \chib_1 \right)^2
\nonumber
\\
&
+   \frac{1}{2} (1 - \Omega_K) \left[  \frac{3}{4}   \phib_1^2 -   \phib_2 
+  \frac{1}{2} \phib_1 \chib_1 \left( 1 - \frac{1}{12} \chib_1 \right)
 + \frac{1}{4} \Omega_K \phib_1^2
\right]
\\
\ldots 
\nonumber
\end{align}
\label{h_FRW_pert}
\end{subequations}
where $\chib_n = d\phib_n/d\ln a$ and  $ \Omega_K = \kappa a^{-2}/ \bar{H}$.
The final solution is constructed from \eqref{phi_FRW_pert} and \eqref{h_FRW_pert} with the help of \eqref{S_op}. In particular one proceeds 
as $\phib_1 \rightarrow h_1 \rightarrow \phib_2 \rightarrow h_2 \rightarrow \dots $ and so forth.

A particular case of interest is a flat universe with $\Omega_K=0$ and matter with constant equation of state $w$. Then
\begin{equation}
\phib =  \phib_0 \left[ 1  
+ 2 (\alpha  + \alpha^2 + \alpha^3)  \ln a 
+( 2 \alpha^2  + 4 \alpha^3) \ln^2 a 
+ \frac{4}{3} \alpha^3 \ln^3 a
+ \ldots
\right]
\label{phi_FRW_lna}
\end{equation}
and
\begin{equation}
H =  \bar{H} \left[ 1
- \frac{1}{6} \frac{5-3w}{1-w} \alpha
-  \frac{(1-3w)(3-w)}{24(1-w)^2} \alpha^2
- \left(\alpha  +  \frac{(1-3w)}{6(1-w)} \alpha^2  \right) \ln a
+   \frac{1}{2} \alpha^2 \ln^2 a  
+ \ldots
\right]  
\label{H_FRW_lna}
\end{equation}
where
\begin{equation}
\alpha =  \frac{1-3w}{1-w} \epsilon
\end{equation}
Clearly, in a radiation dominated Universe, the solution is $\phib = constant$ and $H = \bar{H}$ as we would expect from the fact 
that the scalar couples to the trace of the energy-momentum tensor.

Imposing $w=-1$ in \eqref{phi_FRW_lna} and \eqref{H_FRW_lna} and keeping terms to $\Order(\epsilon)$ gives
\eqref{phi_FRW_lna_Lambda} and \eqref{H_FRW_lna_Lambda} after letting  $\bar{H} = H_0 = \sqrt{\frac{\Lambda}{3\phib_0}}$.

\bibliographystyle{JHEP.bst}
\bibliography{references}

\providecommand{\href}[2]{#2}\begingroup\raggedright\begin{thebibliography}{10}

\bibitem{Perivolaropoulos2008}
L.~Perivolaropoulos, {\it {Six Puzzles for LCDM Cosmology}},
  \href{http://arxiv.org/abs/0811.4684}{{\tt arXiv:0811.4684}}.

\bibitem{FamaeyMcGaugh2011}
B.~Famaey and S.~McGaugh, {\it {Modified Newtonian Dynamics (MOND):
  Observational Phenomenology and Relativistic Extensions}},  {\em Living Rev.
  Rel.} {\bf 15} (2012) 10, [\href{http://arxiv.org/abs/1112.3960}{{\tt
  arXiv:1112.3960}}].

\bibitem{SahniShafielooStarobinsky2014}
V.~Sahni, A.~Shafieloo, and A.~A. Starobinsky, {\it {Model independent evidence
  for dark energy evolution from Baryon Acoustic Oscillations}},  {\em
  Astrophys. J.} {\bf 793} (2014), no.~2 L40,
  [\href{http://arxiv.org/abs/1406.2209}{{\tt arXiv:1406.2209}}].

\bibitem{BullEtAl2015}
P.~Bull et~al., {\it {Beyond $\Lambda$CDM: Problems, solutions, and the road
  ahead}},  {\em Phys. Dark Univ.} {\bf 12} (2016) 56--99,
  [\href{http://arxiv.org/abs/1512.05356}{{\tt arXiv:1512.05356}}].

\bibitem{Weinberg1988}
S.~Weinberg, {\it {The Cosmological Constant Problem}},  {\em Rev. Mod. Phys.}
  {\bf 61} (1989) 1--23.

\bibitem{Martin2012}
J.~Martin, {\it {Everything You Always Wanted To Know About The Cosmological
  Constant Problem (But Were Afraid To Ask)}},  {\em Comptes Rendus Physique}
  {\bf 13} (2012) 566--665, [\href{http://arxiv.org/abs/1205.3365}{{\tt
  arXiv:1205.3365}}].

\bibitem{AntoniadisIliopoulosTomaras1986}
I.~Antoniadis, J.~Iliopoulos, and T.~N. Tomaras, {\it {Quantum Instability of
  De Sitter Space}},  {\em Phys. Rev. Lett.} {\bf 56} (1986) 1319.

\bibitem{FloratosIliopoulosTomaras1987}
E.~G. Floratos, J.~Iliopoulos, and T.~N. Tomaras, {\it {Tree Level Scattering
  Amplitudes in De Sitter Space Diverge}},  {\em Phys. Lett.} {\bf B197} (1987)
  373--378.

\bibitem{Polyakov2012}
A.~M. Polyakov, {\it {Infrared instability of the de Sitter space}},
  \href{http://arxiv.org/abs/1209.4135}{{\tt arXiv:1209.4135}}.

\bibitem{Woodard2014}
R.~P. Woodard, {\it {Perturbative Quantum Gravity Comes of Age}},  {\em Int. J.
  Mod. Phys.} {\bf D23} (2014), no.~09 1430020,
  [\href{http://arxiv.org/abs/1407.4748}{{\tt arXiv:1407.4748}}].

\bibitem{CopelandSamiTsujikawa2006}
E.~J. Copeland, M.~Sami, and S.~Tsujikawa, {\it {Dynamics of dark energy}},
  {\em Int. J. Mod. Phys.} {\bf D15} (2006) 1753--1936,
  [\href{http://arxiv.org/abs/hep-th/0603057}{{\tt hep-th/0603057}}].

\bibitem{CliftonEtAl2012}
T.~Clifton, P.~G. Ferreira, A.~Padilla, and C.~Skordis, {\it {Modified Gravity
  and Cosmology}},  {\em Phys. Rept.} {\bf 513} (2012) 1--189,
  [\href{http://arxiv.org/abs/1106.2476}{{\tt arXiv:1106.2476}}].

\bibitem{Padmanabhan2016}
T.~Padmanabhan, {\it {Do We Really Understand the Cosmos?}},
  \href{http://arxiv.org/abs/1611.03505}{{\tt arXiv:1611.03505}}.

\bibitem{AbellEtAl2009}
{\bf LSST Science, LSST Project} Collaboration, P.~A. Abell et~al., {\it {LSST
  Science Book, Version 2.0}},  \href{http://arxiv.org/abs/0912.0201}{{\tt
  arXiv:0912.0201}}.

\bibitem{LaureijsEtAl2011}
{\bf EUCLID} Collaboration, R.~Laureijs et~al., {\it {Euclid Definition Study
  Report}},  \href{http://arxiv.org/abs/1110.3193}{{\tt arXiv:1110.3193}}.

\bibitem{YahyaEtAl2014}
S.~Yahya, P.~Bull, M.~G. Santos, M.~Silva, R.~Maartens, P.~Okouma, and
  B.~Bassett, {\it {Cosmological performance of SKA HI galaxy surveys}},  {\em
  Mon. Not. Roy. Astron. Soc.} {\bf 450} (2015), no.~3 2251--2260,
  [\href{http://arxiv.org/abs/1412.4700}{{\tt arXiv:1412.4700}}].

\bibitem{SpergelEtAl2015}
D.~Spergel et~al., {\it {Wide-Field InfrarRed Survey Telescope-Astrophysics
  Focused Telescope Assets WFIRST-AFTA 2015 Report}},
  \href{http://arxiv.org/abs/1503.03757}{{\tt arXiv:1503.03757}}.

\bibitem{AghamousaEtAl2016}
{\bf DESI} Collaboration, A.~Aghamousa et~al., {\it {The DESI Experiment Part
  I: Science,Targeting, and Survey Design}},
  \href{http://arxiv.org/abs/1611.00036}{{\tt arXiv:1611.00036}}.

\bibitem{PavlidouTomaras2013}
V.~Pavlidou and T.~N. Tomaras, {\it {Where the world stands still: turnaround
  as a strong test of $\Lambda$ CDM cosmology}},  {\em JCAP} {\bf 1409} (2014)
  020, [\href{http://arxiv.org/abs/1310.1920}{{\tt arXiv:1310.1920}}].

\bibitem{EingornKudinovaZhuk2012}
M.~Eingorn, A.~Kudinova, and A.~Zhuk, {\it {Dynamics of astrophysical objects
  against the cosmological background}},  {\em JCAP} {\bf 1304} (2013) 010,
  [\href{http://arxiv.org/abs/1211.4045}{{\tt arXiv:1211.4045}}].

\bibitem{Stuchlik1983}
Z.~{Stuchlik}, {\it {The motion of test particles in black-hole backgrounds
  with non-zero cosmological constant}},  {\em Bulletin of the Astronomical
  Institutes of Czechoslovakia} {\bf 34} (Mar., 1983) 129--149.

\bibitem{StuchlikHledik1999}
Z.~Stuchlik and S.~Hledik, {\it {Some properties of the Schwarzschild-de Sitter
  and Schwarzschild - anti-de Sitter space-times}},  {\em Phys. Rev.} {\bf D60}
  (1999) 044006.

\bibitem{PavlidouTetradisTomaras2014}
V.~Pavlidou, N.~Tetradis, and T.~N. Tomaras, {\it {Constraining Dark Energy
  through the Stability of Cosmic Structures}},  {\em JCAP} {\bf 1405} (2014)
  017, [\href{http://arxiv.org/abs/1401.3742}{{\tt arXiv:1401.3742}}].

\bibitem{TanoglidisPavlidouTomaras2014}
D.~Tanoglidis, V.~Pavlidou, and T.~Tomaras, {\it {Testing $\Lambda$CDM
  cosmology at turnaround: where to look for violations of the bound?}},  {\em
  JCAP} {\bf 1512} (2015), no.~12 060,
  [\href{http://arxiv.org/abs/1412.6671}{{\tt arXiv:1412.6671}}].

\bibitem{TanoglidisPavlidouTomaras2016}
D.~Tanoglidis, V.~Pavlidou, and T.~Tomaras, {\it {Turnaround overdensity as a
  cosmological observable: the case for a local measurement of $\Lambda$}},
  \href{http://arxiv.org/abs/1601.03740}{{\tt arXiv:1601.03740}}.

\bibitem{Lee2016}
J.~Lee, {\it {On the Universality of the Bound-Zone Peculiar Velocity
  Profile}},  \href{http://arxiv.org/abs/1603.06672}{{\tt arXiv:1603.06672}}.

\bibitem{BhattacharyaDialektopoulosTomaras2015}
S.~Bhattacharya, K.~F. Dialektopoulos, and T.~N. Tomaras, {\it {Large scale
  structures and the cubic galileon model}},  {\em JCAP} {\bf 1605} (2016),
  no.~05 036, [\href{http://arxiv.org/abs/1512.08856}{{\tt arXiv:1512.08856}}].

\bibitem{Faraoni2015}
V.~Faraoni, {\it {Turnaround radius in modified gravity}},  {\em Phys. Dark
  Univ.} {\bf 11} (2016) 11--15, [\href{http://arxiv.org/abs/1508.00475}{{\tt
  arXiv:1508.00475}}].

\bibitem{FaraoniLapierrePrain2015}
V.~Faraoni, M.~Lapierre-Léonard, and A.~Prain, {\it {Turnaround radius in an
  accelerated universe with quasi-local mass}},  {\em JCAP} {\bf 1510} (2015),
  no.~10 013, [\href{http://arxiv.org/abs/1508.01725}{{\tt arXiv:1508.01725}}].

\bibitem{McVittie1933}
G.~C. McVittie, {\it {The mass-particle in an expanding universe}},  {\em Mon.
  Not. Roy. Astron. Soc.} {\bf 93} (1933) 325--339.

\bibitem{Gao2004}
C.~J. Gao, {\it {Arbitrary dimensional Schwarzschild-FRW black holes}},  {\em
  Class. Quant. Grav.} {\bf 21} (2004) 4805--4810,
  [\href{http://arxiv.org/abs/gr-qc/0411033}{{\tt gr-qc/0411033}}].

\bibitem{KaloperKlebanMartin2010}
N.~Kaloper, M.~Kleban, and D.~Martin, {\it {McVittie's Legacy: Black Holes in
  an Expanding Universe}},  {\em Phys. Rev.} {\bf D81} (2010) 104044,
  [\href{http://arxiv.org/abs/1003.4777}{{\tt arXiv:1003.4777}}].

\bibitem{BransDicke1961}
C.~Brans and R.~H. Dicke, {\it {Mach's principle and a relativistic theory of
  gravitation}},  {\em Phys. Rev.} {\bf 124} (1961) 925--935.

\bibitem{BertottiIessTortora2003}
B.~Bertotti, L.~Iess, and P.~Tortora, {\it {A test of general relativity using
  radio links with the Cassini spacecraft}},  {\em Nature} {\bf 425} (2003)
  374.

\bibitem{Will2014}
C.~M. Will, {\it The confrontation between general relativity and experiment},
  {\em Living Reviews in Relativity} {\bf 17} (2014), no.~4.

\bibitem{BakerPsaltisSkordis2015}
T.~Baker, D.~Psaltis, and C.~Skordis, {\it {Linking Tests of Gravity On All
  Scales: from the Strong-Field Regime to Cosmology}},  {\em Astrophys. J.}
  {\bf 802} (2015) 63, [\href{http://arxiv.org/abs/1412.3455}{{\tt
  arXiv:1412.3455}}].

\bibitem{AvilezSkordis2013}
A.~Avilez and C.~Skordis, {\it {Cosmological constraints on Brans-Dicke
  theory}},  {\em Phys. Rev. Lett.} {\bf 113} (2014), no.~1 011101,
  [\href{http://arxiv.org/abs/1303.4330}{{\tt arXiv:1303.4330}}].

\bibitem{Horndeski1974}
G.~W. Horndeski, {\it {Second-order scalar-tensor field equations in a
  four-dimensional space}},  {\em Int.J.Theor.Phys.} {\bf 10} (1974) 363--384.

\bibitem{DeffayetGaoSteer2011}
C.~Deffayet, X.~Gao, D.~A. Steer, and G.~Zahariade, {\it {From k-essence to
  generalised Galileons}},  {\em Phys.Rev.} {\bf D84} (2011) 064039.

\bibitem{Vainshtein1972}
A.~I. Vainshtein, {\it {To the problem of nonvanishing gravitation mass}},
  {\em Phys. Lett.} {\bf B39} (1972) 393--394.

\bibitem{KhouryWeltman2003}
J.~Khoury and A.~Weltman, {\it {Chameleon fields: Awaiting surprises for tests
  of gravity in space}},  {\em Phys. Rev. Lett.} {\bf 93} (2004) 171104,
  [\href{http://arxiv.org/abs/astro-ph/0309300}{{\tt astro-ph/0309300}}].

\bibitem{HinterbichlerKhoury2010}
K.~Hinterbichler and J.~Khoury, {\it {Symmetron Fields: Screening Long-Range
  Forces Through Local Symmetry Restoration}},  {\em Phys. Rev. Lett.} {\bf
  104} (2010) 231301, [\href{http://arxiv.org/abs/1001.4525}{{\tt
  arXiv:1001.4525}}].

\bibitem{LiEtAl2015}
J.-X. Li, F.-Q. Wu, Y.-C. Li, Y.~Gong, and X.-L. Chen, {\it {Cosmological
  constraint on Brans-Dicke Model}},  {\em Res. Astron. Astrophys.} {\bf 15}
  (2015), no.~12 2151--2163, [\href{http://arxiv.org/abs/1511.05280}{{\tt
  arXiv:1511.05280}}].

\bibitem{BallardiniEtAl2016}
M.~Ballardini, F.~Finelli, C.~Umiltà, and D.~Paoletti, {\it {Cosmological
  constraints on induced gravity dark energy models}},  {\em JCAP} {\bf 1605}
  (2016), no.~05 067, [\href{http://arxiv.org/abs/1601.03387}{{\tt
  arXiv:1601.03387}}].

\bibitem{AlonsoEtAl2016}
D.~Alonso, E.~Bellini, P.~G. Ferreira, and M.~Zumalacarregui, {\it {The
  Observational Future of Cosmological Scalar-Tensor Theories}},
  \href{http://arxiv.org/abs/1610.09290}{{\tt arXiv:1610.09290}}.

\bibitem{AvilezSkordisSong2016}
A.~Avilez, C.~Skordis, and Y.-S. Song, {\it Constraining the {B}rans-{D}icke
  theory of gravity with redshift space distortions from future galaxy surveys
  (in preparation)}, .

\bibitem{BhattacharyaEtAl2015}
S.~Bhattacharya, K.~F. Dialektopoulos, A.~E. Romano, and T.~N. Tomaras, {\it
  {Brans-Dicke Theory with $\Lambda>0$: Black Holes and Large Scale
  Structures}},  {\em Phys. Rev. Lett.} {\bf 115} (2015), no.~18 181104,
  [\href{http://arxiv.org/abs/1505.02375}{{\tt arXiv:1505.02375}}].

\bibitem{DeyBhattacharyaSarkar2014}
D.~Dey, K.~Bhattacharya, and T.~Sarkar, {\it {Galactic space-times in modified
  theories of gravity}},  {\em Gen. Rel. Grav.} {\bf 47} (2015) 103,
  [\href{http://arxiv.org/abs/1407.0319}{{\tt arXiv:1407.0319}}].

\bibitem{Eingorn2015}
M.~Eingorn, {\it {First-order Cosmological Perturbations Engendered by
  Point-like Masses}},  {\em Astrophys. J.} {\bf 825} (2016), no.~2 84,
  [\href{http://arxiv.org/abs/1509.03835}{{\tt arXiv:1509.03835}}].

\bibitem{Skordis2016}
C.~Skordis, {\it Conditions for the equivalence of static-spherically symmetric
  to almost friedman-robertson-walker spacetimes (in preparation)}, .

\bibitem{Weinberg1972}
S.~Weinberg, {\em {Gravitation and Cosmology}}.
\newblock John Wiley and Sons, New York, 1972.

\bibitem{Will1981}
C.~Will, {\em {Theory and experiment in gravitational physics}}.
\newblock Cambridge University Press, Cambridge, 1981.

\bibitem{Padmanabhan1993}
T.~Padmanabhan, {\em {Structure Formation in the Universe}}.
\newblock Cambridge University Press, Cambridge, 1993.

\end{thebibliography}\endgroup

\end{document}